\documentclass[12pt]{article}
\begin{document}
\title{DISCRETE SPACE TIME AND DARK ENERGY}
\author{B.G. Sidharth\\
International Institute for Applicable Mathematics \& Information Sciences\\
Hyderabad (India) \& Udine (Italy)\\
B.M. Birla Science Centre, Adarsh Nagar, Hyderabad - 500 063 (India)}
\date{}
\maketitle
\begin{abstract}
In recent times, Discrete Space Time Architectures are being considered, in the context of Quantum Gravity, Quantum Super Strings, Dark Energy and so on. We show that such a scheme is intimately tied up with a varying $G$ cosmology, which again explains otherwise inexplicable observations like the anomalous accelerations of Pioneer space crafts as also considerations involving the Zero Point Field, Random Electrodynamics and the derivation of Quantum Mechanical effects therefrom as also intertial mass.
\end{abstract}
\section{Introduction}
Quantum Gravity and Quantum Super String Theory work at a minimum spacetime interval. This is at the Planck scale, and is a major departure from earlier theories, be they Classical or Quantum, where a differenciable spacetime manifold is considered \cite{garay,kempf,mup,nc117}.\\
According to 't Hooft \cite{hooft}, space time has to be discrete. Infact, in his words, ``...It is somewhat puzzling... why the lattice structure of space and time had escaped attention from other investigators...'' Similarly the author had, working independantly on these lines, developed a model that correctly predicted dark energy and an accelerating expanding universe, all of which was subsequently confirmed by observation \cite{ijmpa,ijtp,cu,science}.
\section{Discrete Space Time Gravitation}
We will introduce the minimum cut off into the Schwarzschild geometry and show that this is equivalent to a scheme in which the gravitational constant slowly varies with time.\\
Our starting point is the Schwarzschild geometry \cite{mwt}
\begin{equation}
d \tau^2_0 = -\left(1 - \frac{2GM}{r}\right) dt^2 + \frac{dr^2}{1-2GM/r} + r^2 (d\Theta^2 + sin^2 \Theta d\phi^2)\label{e1}
\end{equation}
In the above we consider units in which $c = 1$.\\
Let us now replace 
$$\Omega = 1-\frac{2MG}{r}$$
by
$$\Omega = 1-\frac{2MG}{r-l} = 1-\frac{2MG}{r} \left\{ 1 + \frac{l}{r}\right\}, r > > l.$$
Effectively we are removing the singularity of the Schwarzschild metric. As
$$\Omega^{-1} = 1+\frac{2MG}{r} \left\{ 1+\frac{l}{r}\right\},$$
we have from (\ref{e1}) for large $r$
$$
d\tau^2 - d\tau^2_0 = \frac{2MGl}{r^2} dt^2 - \frac{2MGl}{r^2} dr^2$$
\begin{equation}
= \frac{2MG}{r} (\frac{l}{r}) [dt^2 - dr^2]\label{e2}
\end{equation}
In (\ref{e2}) $d \tau^2$ is the modified metric and $d \tau^2_0$ is given by (\ref{e1}).\\
Whence we have
\begin{equation}
d\tau^2 = d\tau^2_0 - \frac{2MG}{r}(\frac{l}{r}) (dt^2 - dr^2)\label{e3}
\end{equation}
From (\ref{e3}) we can see that
\begin{equation}
g_{00} = 1 - \frac{2MG}{r} (1 + \frac{l}{r})\label{e4}
\end{equation}
Equation (\ref{e4}) shows that there is an extra force
$$\propto \frac{GMl}{r^3}$$
Alternatively (\ref{e4}) is equivalent to $G$ acquiring a multiplying factor which is very nearly unity. We could also obtain this as follows:
\begin{equation}
G(t-\tau) = G + \frac{\tau G}{t} = G(1+\beta), \beta  = \tau/t\label{e5},
\end{equation}
if we require, 
\begin{equation}
\dot {G} = -G/t\label{e6}
\end{equation}
It may be mentioned that (\ref{e6}) is consistent with observation \cite{norman,melnikov}.
(\ref{e5}) immediately gives (\ref{e4}) if we have,
\begin{equation}
\frac{r}{t} = \frac{l}{\tau}\label{e7}
\end{equation}
We can immediately see that (\ref{e7}) is correct if $l$ is a Compton length and $\tau$ the corresponding Compton time and $t$ and $r$ are the age and radius of the universe (remembering that $c = 1$). The Planck scale infact is the Compton scale for a Planck mass. In any case we have with discrete spacetime, the equation (\ref{e6}).\\
Time varying $G$ cosmologies have been considered though in slightly different contexts (Cf.for example references \cite{narlikar,barrow,cu}). Infact apart from several well known observations, (\ref{e6}) also explains less well known effects like the decrease in the orbital period and diameter of binary pulsars or the mysterious anomalous accelerations of the Pioneer space crafts which have defied any other explanation so far \cite{nc115,ffp5,and,and2}.
\section{Binary Pulsars}
Cosmologies with a variable $G$ as mentioned above have been considered in the past \cite{narlikar,barrow} and more recently by the author, though in a slightly different context \cite{ijmpa,ijtp,cu}. Essentially in this latter case, there is a background dark energy which leads to an accelerating ever expanding universe (while Gravitation manifests itself in the spirit of Sakharov's theory). This prediction was confirmed observationally in 1998 and dark energy itself was finally confirmed in 2003 and declared as the breakthrough of the year \cite{science}.\\
It has been shown that this scheme explains many observational results like the precesion of the perihelion, the bending of light and so on \cite{nc115,csf}. We would now like to point out that other delicate effects like the observed decrease in orbital parameters of a binary pulsar, an anomalous acceleration for the planets, and an as yet unexplained anomalous acceleration of the Pioneer space crafts can also be explained within this scheme.\\
Our starting point is the equation
\begin{equation}
G \sim \frac{G_0}{T}\label{e1a}
\end{equation}
where $T$ is the age of the universe. It may be noted that this is in agreement with observational limits \cite{bgsfpl,norman}.\\
Let us now consider an object revolving about another object, as in the case of the binary pulsar \cite{gold}. The gravitational energy of the system is  given by,
\begin{equation}
\frac{GMm}{L} = const.\label{e2a}
\end{equation}
Whence
\begin{equation}
\frac{\mu}{L} \equiv \frac{GM}{L} = const.\label{e3a}
\end{equation}
For variable $G$ we have, after the elapse of time $t$,
\begin{equation}
\mu = \mu_0 - tK\label{e4a}
\end{equation}
where
\begin{equation}
K \equiv \dot {\mu}\label{e5a}
\end{equation}
We take $\dot {\mu}$ to be a constant, in view of the fact that $G$ varies very sowly, as can be seen from (\ref{e1a}).\\
Using (\ref{e3a}) and (\ref{e4a}), we get
$$L = L_0(1 - \alpha K)$$
Where
\begin{equation}
\alpha  = \frac{t}{\mu_0}\label{e6a}
\end{equation}
Let $t$ to be the period of revolution. Using (\ref{e6a}) we get,
\begin{equation}
\delta L = \frac{LtK}{\mu_0}\label{e7a}
\end{equation}
We also know from standard theory, (Cf.ref.\cite{gold})
\begin{equation}
t = \frac{2\pi}{h} L^2 = \frac{2\pi}{\sqrt{\mu}}\label{e8a}
\end{equation}
\begin{equation}
t^2 = \frac{4\pi^2L^3}{\mu}\label{e9a}
\end{equation}
Using (\ref{e7a}), (\ref{e8a}) and (\ref{e9a}) a little manipulation gives
\begin{equation}
\delta t = \frac{2t^2K}{\mu_0}\label{e10a}
\end{equation}
(\ref{e7a}) and (\ref{e10a}) show that there is a decrease in the size of the orbit, as also in the orbital period. Such a decrease in the orbital period has been observed in the case of binary pulsars \cite{ohan,will}. Let us consider the binary pulsar PSR $1913 + 16$ observed by Taylor and co-workers (Cf.ref.\cite{will}). In this case it is known that, $t$ is $8$ hours while $v$, the orbital speed is $3 \times 10^7 cms$ per second. It is easy to calculate from the above
$$\mu_0 = 10^4 \times v^3 \sim 10^{26}$$
which gives $M \sim 10^{33}gms$, which of course agrees with observation and is a check. On using (\ref{e1a}) and (\ref{e5a}), we get
\begin{equation}
\Delta t = \eta \times 10^{-5} sec/yr,\, \eta \approx 8\label{e11a}
\end{equation}
Indeed (\ref{e11a}) is in good agreement with the carefully observed value of $\eta \approx 7.5$ (Cf.refs.\cite{ohan,will}).\\
It may be mentioned that this same effect has been interpreted as being due to gravitational radiation, even though there are some objections to the calculation in this case (Cf.ref.\cite{will}).
\section{Anomalous Accelerations}
We next consider observed but inexplicable anomalous accelerations in the Solar System. Our starting point is the well known equation for orbits
\begin{equation}
\frac{\lambda}{r} = (1+e cos \Theta ), \quad \lambda = \frac{l^2}{GM}, l = r^2 \dot \Theta , r\dot \Theta = v,\label{e12a}
\end{equation}
From (\ref{e12}) it follows that
$$
\lambda = \frac{r^4\dot \Theta^2}{GM} = \frac{r^2v^2}{GM} = r (1+e cos \Theta )$$
Whence we get
\begin{equation}
v^2 = \frac{GM}{r} (1 + e cos \Theta )\label{e13a}
\end{equation}
Differenciating (\ref{e13a}), and using (\ref{e1a}), we have
\begin{equation}
v \dot {v} \approx -\frac{GM}{2Tr} (1 + e cos \Theta )-\frac{GM}{r^2} \dot {r}(1+e cos \Theta )\label{e14a}
\end{equation}
It must be mentioned that the first term on the right side of equation (\ref{e14a})is the varying $G$ effect unlike the second term which appears in the usual theory. (\ref{e14a}) shows that there is an anomalous acceleration given by 
\begin{equation}
a_r = \langle \dot v \rangle_{\mbox{anom}} = \frac{-GM}{2T r v} (1+e cos \Theta )\label{e15a}
\end{equation}
Further from (\ref{e12a}) we have on differenciation
$$\dot r (1 + e cos \Theta ) - \frac{e sin \Theta l}{r} = \dot {\lambda} = -\frac{l^2}{GMT}$$
Whence we get
$$\dot r (1+e cos \Theta ) \approx - \frac{\lambda}{T} \sim -1$$
If we take the arbitrary polar axis such that at the given moment of observation, $\Theta = 0$, this gives, as $\dot {r} \approx v$ in this case,
\begin{equation}
v (1 + e) \sim 1\label{e16a}
\end{equation}
Using (\ref{e16a}) in (\ref{e15a}) and also using (\ref{e12a}), we get
$$a_r \sim -\frac{GM}{2 T rv} (1 + e) \approx - \frac{GM}{2Tr}(1+e)^2$$
Whence
\begin{equation}
a_r \approx -\frac{GM}{2T\lambda} (1+e)^3\label{e17a}
\end{equation}
(\ref{e17a}) shows that there is a constant inward anomalous acceleration. If we use the data for the Pioneer space crafts \cite{and} in (\ref{e17a}) we get \cite{ffp5} 
\begin{equation}
a_r \sim -10^{-7} cm/sec^2\label{e18a}
\end{equation}
(\ref{e18a}) has been observed by Anderson and co-workers for the past several years and all their attempts to explain this anomalous acceleration have failed \cite{and2}.\\
Similarly in the case of the earth (and other planets) we can deduce an anomalous inward radial acceleration $\sim 10^{-9}cms/sec^2$ which is known to be the case \cite{kuhne}.
\section{Dark Energy Considerations}
We note that as is well known, such a background ZPF can explain the Quantum Mechanical spin half as also the anomalous $g = 2$ factor for an otherwise purely classical electron \cite{sachi,boyer}. The key point here is (Cf.ref.\cite{sachi}) that the classical angular momentum $\vec r \times m \vec v$ does not satisfy the Quantum Mechanical  commutation rule for the angular momentum $\vec J$. However when we introduce the background Zero Point Field, the momentum now becomes
\begin{equation}
\vec J = \vec r \times m \vec v + (e/2c) \vec r \times (\vec B \times \vec r) + (e/c) \vec r \times \vec A^0 ,\label{e5b}
\end{equation}
where $\vec A^0$ is the vector potential associated with the ZPF and $\vec B$ is an external magnetic field introduced merely for convenience, and which can be made vanishingly small.\\
It can be shown that $\vec J$ in (\ref{e5b}) satisfies the Quantum Mechanical commutation relation for $\vec J \times \vec J$. At the same time we can deduce from (\ref{e5b})
\begin{equation}
\langle J_z \rangle = - \frac{1}{2} \hbar \omega_0/|\omega_0|\label{e6b}
\end{equation}
Relation (\ref{e6b}) gives the correct Quantum Mechanical results referred to above.\\
From (\ref{e5b}) we can also deduce that
\begin{equation}
l = \langle r^2 \rangle^{\frac{1}{2}} = \left(\frac{\hbar}{mc}\right)\label{e7b}
\end{equation}
(\ref{e7b}) shows that the mean dimension of the region in which the fluctuation contributes is of the order of the Compton wavelength of the electron.\\
As a further confirmation, we can similarly deduce that there is a mean time interval of the order of the electrons' Compton time. For this we note that the energy of the ZPF is given by \cite{boyer}
$$\frac{1}{8\pi}\langle E^2 + B^2 \rangle = \frac{1}{8\pi} \sum^{2}_{\lambda = 1} \int d^2 kh^2 (\omega_\hbar)$$
\begin{equation}
= \int^\infty_{k=0} dk k^2h^2(\omega_\hbar) = \int^\infty_{\omega = 0} d\omega^{\omega^2}_{c^3}-h^2(\omega),\label{e8b}
\end{equation}
where the spectral energy density is given by
\begin{equation}
\rho (\omega) =  \frac{\hbar}{2\pi^2} \frac{\omega^3}{c^3}\label{e9b}
\end{equation}
If now we denote the impulsive change of momentum in the interval $\tau$ due to the buffetting of the particle by the ZPF by 
$$\langle \Delta^2 \rangle^{\frac{1}{2}}$$
then we can easily deduce that by (\ref{e8b}) and (\ref{e9b}), 
\begin{equation}
\langle \Delta^2 \rangle = \left(\Gamma \pi^4 c^4 \tau/5\omega^2 \right) \cdot \rho^2 (\omega ,T)\label{e10b}
\end{equation}
where
\begin{equation}
\rho (\omega , T) = \left(\frac{\omega^2}{2\pi^2c^3}\right) \langle m v^2 \rangle\label{e11b}
\end{equation}
In (\ref{e11b}) $v$ is the root mean square velocity of the rapidly vibrating particle, which we take to be $c$, the velocity of light. Then we can deduce from (\ref{e10b}), on using the fact that the magnitude of the impulsive change of the momentum would be $mc$, that
\begin{equation}
\tau \sim \frac{\hbar}{mc^2}\label{e12b}
\end{equation}
(\ref{e12b}) shows that $\tau$ is of the order of the Compton time as is to be expected from (\ref{e7b}).\\
We can interpret the above in the following manner. Let us in (\ref{e11b}) take the velocity $v$ to be the velocity of light. This automatically means that the mass $m$ is vanishingly small. Then as a result of the ZPF we can deduce from (\ref{e7b}), (\ref{e9b}), (\ref{e10b}), (\ref{e11b}) and (\ref{e12b}) that the energy of the particle is $\frac{r \Delta}{\tau}$ which is $mc^2$. In other words the otherwise massless particles acquires its inertial mass $m$ as a result of the ZPF at the Compton scale. Indeed this has been shown earlier also \cite{ijmpa}. Interestingly, in the context of (\ref{e11b}) and the ZPF, it can be easily shown that the thermal wavelength, in the limit $v \to c$ becomes the Compton wavelength (Cf.\cite{ijmpa}). We can see all this briefly as follows:\\
Starting with the energy of the ZPF
$$B^2 \sim \frac{\hbar c}{\lambda^4},$$
where $\lambda$ is the Compton wavelength, we can see that the energy in a volume $\sim \lambda^3$ is given by $mc^2$.\\
From a different perspective, the fluctuation in curvature $R$ over the length $\lambda$ is given by
$$\Delta R \sim \frac{L^*}{\lambda^3},$$
where $L$ is the Planck length. It follows that at the Compton scale of elementary particles
$$\Delta R \sim 1$$
From yet another perspective, if we consider a massless particle undergoing rapid motion in a small volume $V$, the thermal wavelength is given by
$$\frac{V}{N} \sim \lambda_{thermal} \approx \frac{\hbar}{\sqrt{m^2c^2}} = \frac{\hbar}{mc},$$
that is the entire mass of the particle can be recovered from the rapid Zitterbewegung type motions at the Compton scale, as we saw above.\\  
There is another way to see this. In the Santamato deBroglie-Bohm approach it turns out that there is a scalar curvature $R$ in terms of which the Quantum potential $Q$ is given by \cite{santa,afdb}
\begin{equation}
\gamma \frac{\hbar^2}{2m} \dot R = Q, \gamma = \frac{1}{c}\label{e13b}
\end{equation}
In earlier work \cite{cu}, the Quantum potential $Q$ has been shown to equal the energy $mc^2$, at the Compton scale. So we have, at this scale from (\ref{e13b})
$$\Delta R = \frac{2m^2c^2}{\hbar^2\gamma}  \tau = \frac{1}{r}$$
This shows that the curvature is that of a sphere of dimension of the Compton wavelength. In other words the ZPF curls up or condenses the Compton scale to form an elementary particle.\\
We now observe from (\ref{e5b}) that, given the ZPF, in the above context, we have
$$p^\mu \to p^\mu + \frac{e}{c} A^\mu$$
This well known minimum coupling result appears in the Weyl formulation of guage invariant geometry and has been shown to be equivalent to an underlying non commutative geometry arising due to minimum space time intervals which again has been shown to be symptomatic of the double connectivity of the Dirac bispinor wave equation \cite{afdb,nc115}. The minimum space time cut off at the Compton scale too has been shown to follow in the above formulation (\ref{e7b}) and (\ref{e12b}). Thus, the geometry of fermions with the above characteristics is symptomatic of the ZPF at the Compton scale.\\
Finally, it may be seen that the minimum space time interval considerations which lead to (\ref{e6}), can equally well be thought of, equivalently, to $\sqrt{N}$ particles (pions) being created in the Compton time fuzzy interval thus adding to the mass of the universe by the amount $M/\sqrt{N}$. This is because as can be seen, from equation (\ref{e7}), 
$$\frac{r}{l} \approx \frac{t}{\tau} \approx \sqrt{N}.$$ 

\end{document}